\documentclass[superscriptaddress, aps, prl,twocolumn]{revtex4-2}
\usepackage{amsmath,amssymb}
\usepackage{graphicx,hyperref,color}
\usepackage{bm}

\begin{document}

\title{Alignment-Induced Self-Organization of Autonomously Steering Microswimmers: Turbulence, Vortices, and Jets}
\author{Segun Goh}
\affiliation{Theoretical Physics of Living Matter, Institute for Advanced Simulation, Forschungszentrum J{\"u}lich, 52425 J{\"u}lich, Germany}
\author{Elmar Westphal}
\affiliation{Peter Gr\"unberg Institute and J\"ulich Centre for Neutron Science, Forschungszentrum J\"ulich, 52425 J\"ulich, Germany}
\author{Roland G. Winkler}
\affiliation{Theoretical Physics of Living Matter, Institute for Advanced Simulation, Forschungszentrum J{\"u}lich, 52425 J{\"u}lich, Germany}
\author{Gerhard Gompper}
\email{g.gompper@fz-juelich.de}
\affiliation{Theoretical Physics of Living Matter, Institute for Advanced Simulation, Forschungszentrum J{\"u}lich, 52425 J{\"u}lich, Germany}
\date{\today}
\begin{abstract}
Systems of motile microorganisms exhibit a multitude of collective phenomena, including motility-induced phase separation and turbulence.
Sensing of the environment and adaptation of movement plays an essential role in the emergent behavior. We study the collective motion of
wet self-steering polar microswimmers, which align their propulsion direction hydrodynamically with that of their neighbors, by mesoscale
hydrodynamics simulations. The simulations of the employed squirmer model reveal a distinct dependence on the swimmer flow field,
i.e., pullers versus pushers. The collective motion of pushers is characterized by active turbulence, with nearly homogeneous density and a
Gaussian velocity distribution. Pullers exhibit a strong tendency for clustering and display velocity and vorticity distributions with fat exponential
tails; their dynamics is chaotic, with a temporal appearance of vortex rings and fluid jets.
Our results show that the collective behavior of intelligent microswimmers is very diverse and still offers many surprises to be discovered.
\end{abstract}
\maketitle
\section{Introduction}
Emergence of dynamic structures and patterns is an essential feature of biological active motile systems. Examples include microbial swarms~\cite{pedl:92,beer:19}
on the cellular level, to schools of fish~\cite{katz:11,guat:12}, flocks of birds~\cite{ball:08.1,bial:12.1}, 
and collective motion in human crowds \cite{helb:00} on a macroscopic level. Also, in artificial active systems 
consisting of synthetic self-propelled particles and microrobots, the collective dynamics of constituent objects 
is of fundamental importance for their application in engineering and medicine to achieve a large spectrum of functionalities~\cite{xie:19,qian:21,chen:22,xian:23}.

A fundamental aspect in such systems is the active and autonomous motion of the constituting particles. While activity and self-propulsion can give rise to several novel types of collective behaviors, such as motility-induced 
phase separation (MIPS)~\cite{fily:12,butt:13} and active turbulence~\cite{wens:12,qi:22,aran:22}, 
the fact that biological microswimmers are not only motile, but also gather information about their environment 
and adapt their motion by self-steering, remains largely unexplored and yet to be elucidated~\cite{ziep:22,sawi:23,negi:24}. 

Many living organisms are immersed in a fluid medium, and their collective behavior is strongly affected or 
even dominated by hydrodynamics~\cite{koch:11,elge:15}. The hydrodynamic environment is not just the background medium 
in which aquatic microorganisms are based, but it is rather essential for locomotion on the individual 
level as well as inter-organism interactions~\cite{pedl:92}. On the mesoscale, life has adapted to low-Reynolds 
hydrodynamics, e.g., in the emergence of bacterial turbulence~\cite{domb:04,dres:11,aran:22} and in coordinated cell 
migration during embryogenesis~\cite{dene:19,hern:23}; similar physical laws govern the swarming microrobots~\cite{xie:19,cero:23}. However, studying large-scale hydrodynamic systems is challenging as 
the fluid adds a large number of degrees of freedom, while the consideration of large-scale systems is unavoidable to accurately capture long-range hydrodynamics interactions and to minimize potential finite-size effects. So far, only limited studies have been conducted in this direction, particularly in three dimensions.

The goal of our current endeavor is to unravel the emergent collective behavior of systems, which combine two 
essential components of living and artificial active systems, self-steering and hydrodynamics. In the context 
of active matter, the Vicsek model~\cite{vics:95,chat:20} is among the earliest and simplest models for the 
collective directional motion of self-propulsion and self-steering particles due to mutual alignment. While 
the role of alignment interactions has been extensively investigated for dry active systems~\cite{chat:20}, 
hydrodynamic interactions themselves have rather been viewed as a physical alignment mechanism in wet 
systems \cite{rein:18,liu:21a}. However, hydrodynamic propulsion can in fact also destabilize polar order~\cite{simh:02,llop:10,clop:20}, which implies the necessity for an additional stabilization mechanism for alignment.
To achieve stable alignment, we consider a hydrodynamic extension of the Vicsek model. Our active agents, modeled as squirmers~\cite{ligh:52,blak:71,ishi:06,goet:10,thee:16.1}, sense the propulsion direction of neighboring agents and adapt their propulsion direction accordingly by hydrodynamic self-steering \cite{pak:14,goh:23}, with a ``slow" temporal response due to limited maneuverability. We perform large-scale simulations in three-dimensional systems, capturing the fluid environment by the multiparticle collision dynamics (MPC) technique \cite{male:99,kapr:08,gomp:09}, a particle-resolved mesoscale hydrodynamic simulation approach.

\begin{figure*}
    \centering  
    \includegraphics[width= 2\columnwidth]{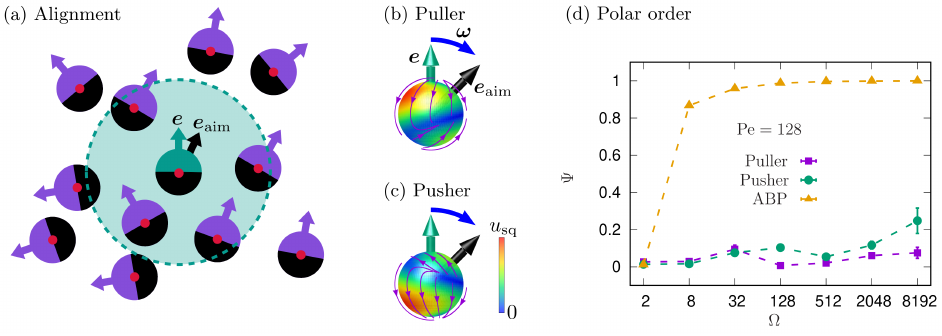}
    \caption{\label{fig:polar}
    {\bf Illustration of the alignment interaction.}  (a) The microswimmer with the orientation ${\bm e}$ (petrol) 
    senses the orientations of neighboring microswimmers (purple) within the sensing range $R_a$ 
    (green dashed circle) and reorients toward ${\bm e}_{\rm aim}$ (black arrow), which is the average 
    orientation of the neighbors determined via Eq.~\eqref{eq:aim}.
    (b) and (c) Non-axisymmetric surface flow fields for hydrodynamic self-steering of (b) puller- and 
    (c) pusher-type microswimmers with angular velocity $\boldsymbol{\omega}$. Adapted from Ref.~\cite{goh:23}.
    (d) Global polar order parameters $\Psi$ for pullers (purple squares), pushers (green circles), and ABPs 
    (yellow triangles) as a function of the maneuverability $\Omega$. Here, ${\rm Pe}=128$.}
\end{figure*}

We observe and characterize the emergence of swarming dynamics in polar active fluids, consisting either of pusher or
puller microswimmers. Our results show that hydrodynamic interactions destabilize polar order, giving rise to 
rich collective spatio-temporal behavior beyond the simple symmetry breaking of the dry Vicsek model. Pusher 
systems feature active turbulence with non-universal scaling exponents in the kinetic energy spectrum, revealing 
a route toward active turbulence via self-steering. Systems of pullers are accompanied by formation of 
dense, swarming clusters driven by hydrodynamic interactions. In particular, formation of toroidal 
structures are observed in the vorticity field, which are characterized by enhanced spatial vorticity-velocity 
cross correlations at short distances. This demonstrates that the formation of vortex rings is a direct 
consequence of strong active jets caused by propulsion and alignment.

\section{Result}

\subsection{Hydrodynamic Vicsek model and Polar Order}

We consider a system of $N$ spherical active microswimmers with radius $R_{\rm sq}$ and instantaneous
orientation ${\bm e}_i$, $i\in\{1, \dots, N\}$. For self-propulsion 
and self-steering, the squirmer model is employed, where the self-propulsion is achieved via an axisymmetric 
surface-slip boundary condition, characterized by the speed $v_0$ and the active stress $\beta$
[see Materials and Methods, Eq.~\eqref{eq:uph}]. Self-steering of microswimmers is modeled via an adaptive non-axisymmetric 
surface-slip boundary condition [see Materials and Methods, Eqs.~\eqref{eq:uph} and~\eqref{eq:uth}), and Fig.~\ref{fig:polar}(b),(c)], which enables the 
squirmer to rotate and consequently reorient toward a desired direction ${\bm e}_{{\rm aim},i}$ with the 
limited angular velocity~\cite{goh:22}
\begin{align} \label{eq:omegai}
{\bm \omega}_i = C_0 {\bm e}_i \times {\bm e}_{{\rm aim},i},
\end{align}
where $C_0$ characterizes the strength of adaptation. 
Accordingly, we introduce two dimensionless parameters, the P\'eclet number ${\rm Pe}$ and the maneuverability 
$\Omega$ in the form
\begin{align}
{\rm Pe} = \frac{v_0}{\sigma D_R}, \quad \Omega = \frac{C_0}{D_R} \, ,
\end{align}
where $\sigma=2R_{\rm sq}$ and $D_R$ are the diameter and the (thermal) rotational diffusion coefficient of a 
squirmer, respectively.

\begin{figure*}
\centering
    \includegraphics[width= 2\columnwidth]{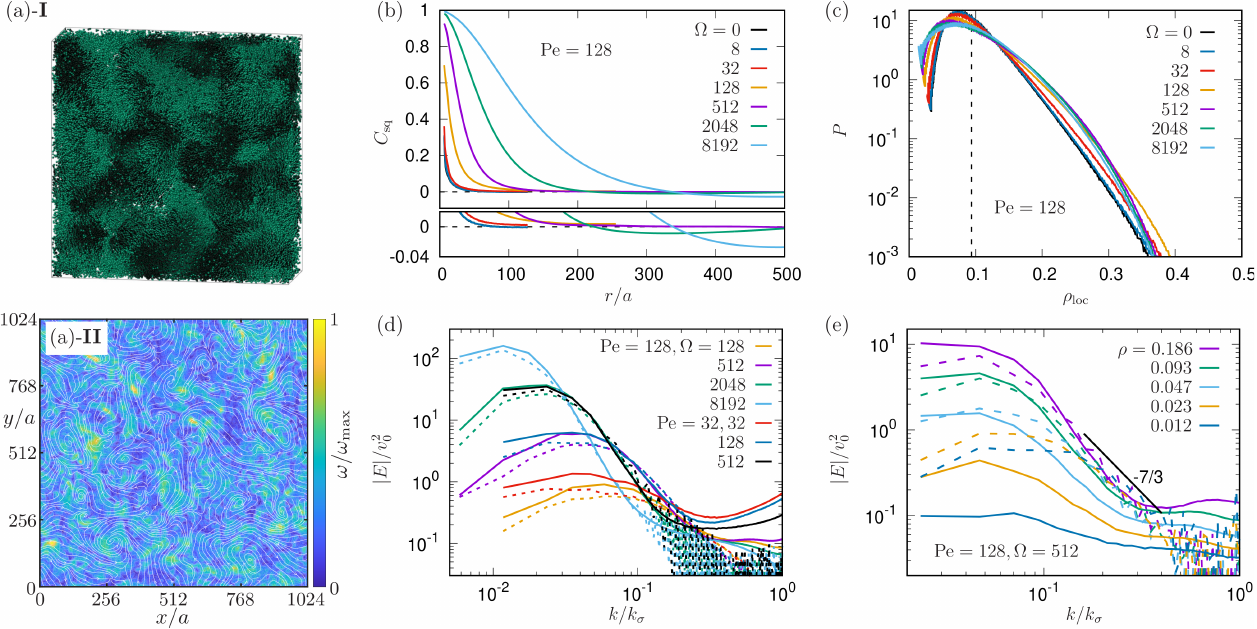}
    \caption{\label{fig:pusher}
    {\bf Active turbulence in pusher systems.}
    (a)-{\bf I} Snapshot of squirmers for ${\rm Pe}=128$ and $\Omega = 2048$. The orientation 
    of each particle is indicated by a petrol hemisphere.
    (a)-{\bf II} Two-dimensional projection of the streamlines of the velocity field (white lines) and the 
    magnitude of the vorticity field (heat map) for ${\rm Pe}=128$ and $\Omega = 2048$.
    (b) Equal-time spatial velocity correlation function as a function of the squirmer distance for various $\Omega$ as indicated in the legend. An enlarged view for small $C_{\rm sq}$ is shown in the bottom panel.
    (c) Squirmer density distribution  as a function of the local packing fraction for various $\Omega$. The global packing fraction is represented by a black dashed line at $\rho_{\rm loc} \approx 0.093$.
    (d) Energy spectrum as a function of the wave number $k$ for $\Omega / {\rm Pe}=1$, $4$, $16$, and $64$, and ${\rm Pe}=32$, $128$, and $512$. $k_{\sigma}=2\pi/\sigma$ is the wave number for the squirmer diameter. 
    (e) Energy spectrum as a function of the wave number for ${\rm Pe} = 128$, $\Omega = 512$, and various densities (legend); here, $L/a=256$. In (d) and (e), solid and dashed lines represent energy spectra of fluids and squirmers, respectively.
}
\end{figure*}

In Eq.~\eqref{eq:omegai}, sensed information of each particle about the orientation of other neighboring particles 
is represented by a vector ${\bm e}_{{\rm aim},i}(\{{\bf e}_j\})$. 
In the interest of the investigation and understanding of the generic collective behavior of an active polar fluid,
we focus here on a ``minimal'' model, where only the swimming and steering mechanism via the adaptive surface flow
field is explicitly considered, whereas sensing and information processing is taken into account implicitly by 
the sensed-information vector ${\bm e}_{\rm aim}$ in Eq.~\eqref{eq:omegai}, see Ref.~\cite{goh:23} for a more 
detailed discussion. As a representative example of information exchanges between intelligent microswimmers, 
we consider a Vicsek-type alignment interaction, where each microswimmer aims at adapting its orientation and
propulsion direction to the average orientation of neighboring particles, see Fig.~\ref{fig:polar}(a) for illustration. Specifically, we employ a non-additive 
rule of orientation adaptation, which results in non-reciprocal interactions between microswimmers, where \cite{chep:21}
\begin{align} \label{eq:aim}
{\bm e}_{{\rm aim},i} = \frac{1}{N_i} \sum_{j \in \Gamma_i} {\bm e}_j.
\end{align}
Here, $\Gamma_i$ is the set of neighbors of the $i$-th particle in its alignment range $R_a$, and $N_i$ is the number 
of neighbors. As apparent from the definition in Eq.~\eqref{eq:aim}, ${\bm e}_{\rm aim}$, which serves as an input 
signal triggering adaptive surface flows according to Eqs.~\eqref{eq:uph} and~\eqref{eq:uth}, is typically not a unit 
vector. Therefore, the magnitude of the adaptation force depends on the strength of orientational order of the 
neighboring microswimmers. We emphasize that in our approach the steering is achieved solely via the modification 
of the surface flow fields, mimicking the autonomous behaviors of microorganisms, in contrast to external 
driving forces, see, e.g., Ref.~\cite{bera:22}. This difference leads to very different behaviors in wet and
dry systems.

For the MPC fluid dynamics simulations, 
we consider a MPC variant with angular momentum conservation~\cite{nogu:08}, see Supplementary Materials (SM), Sec.~S-I for more details.
Our highly parallelized, GPU-accelerated implementation employed for the simulations is based on the framework proposed in Ref.~\cite{west:24}.
 For an accurate 
characterization of emergent behaviors, we consider large system sizes up to $L/a = 1024$, 
where $L$ is the length of the cubic simulation box and $a$ is the side length of a MPC collision cell, and up to $N=884,736$ squirmers. For the squirmers, we choose a sensing range $R_a = 4 R_{\rm sq}$, and a strength of the active stress $\beta = -3$ and $\beta =3$ for 
pushers and pullers, respectively. For most simulations, we consider the packing fraction 
$\rho \equiv (4\pi R_{\rm sq}^3/3)N/L^3 =0.093$ (based on squirmer radius), or 
$\rho_a \equiv (4\pi R_{\rm a}^3/3)N/L^3 \approx 6.0$ (based on sensing range), if not explicitly stated 
otherwise.  For comparison, we also perform simulations of a dry system of aligning self-steering 
``intelligent" active Brownian particles (iABPs) of the same packing fraction.

Results for the global polar order parameter $\Psi \equiv |\sum_i {\bm e}_i |/N$ are shown in 
Fig.~\ref{fig:polar}(d). While the systems are disordered for small maneuverability $\Omega$ in both dry 
and wet cases, a global polar order emerges only in dry systems for large $\Omega$, indicating that 
hydrodynamic interactions destabilize the polar order. Instead of simple polar ordering, our squirmer 
systems show swarming dynamics with and without density modulation depending on the swimming mechanism 
(pusher or puller).

\begin{figure}
	\centering
  	\includegraphics[width= 0.8\columnwidth]{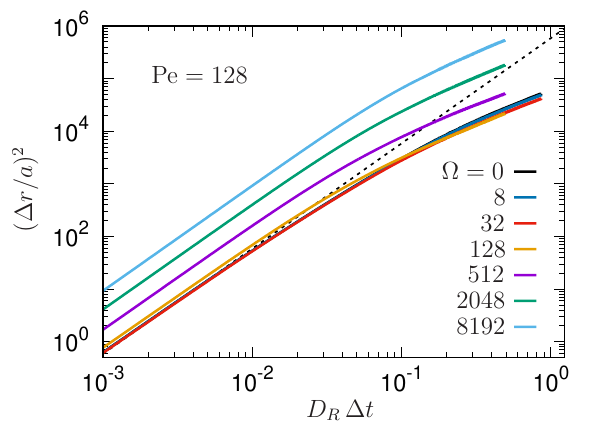}
	\caption{\label{fig:MSD}
	{\bf Mean-square displacement of pushers.} The results are shown as a function of time for various $\Omega$ (legend). The black dashed line indicates 
    the ballistic dynamics of non-interacting squirmers, $(\Delta r)^2 = {v_0}^2 (\Delta t)^2$, and $D_R$ is the rotational diffusion coefficient.
}
\end{figure}

\subsection{Pushers: Active Turbulence via Self-Steering}

Systems of rearly propelled squirmers (pushers) exhibit active turbulence, see Fig.~\ref{fig:pusher}(a)-{\bf I} for a snapshot (also Movie S1), and their collective motion features vortical flows. 
Figure~\ref{fig:pusher}(a)-{\bf II} displays the fluid velocity field reflecting vortical structures and 
fluctuations in the magnitude of the vorticity. Accordingly, equal-time spatial velocity correlations 
become negative at large distances (less than $L/2$) for large $\Omega$, see Fig.~\ref{fig:pusher}(b). 
However, no pronounced density fluctuations are visible, see Fig.~\ref{fig:pusher}(a)-{\bf I}. This is
confirmed by the analysis of the local density distribution via from Voronoi tessellation~\cite{rycr:09}. As shown in Fig.~\ref{fig:pusher}(c), the distribution exhibits a peak near the global squirmer density independent of $\Omega$.

To characterize the dynamics, we examine the kinetic energy spectrum as a representative indicator 
of active turbulence~\cite{qi:22}. We determine the energy spectra for both the squirmer and the fluid. 
For the squirmers, we first calculate spatial velocity correlation functions and then perform a Fourier transform to obtain the energy spectrum. For the fluid, we calculate the energy spectrum directly from the (Eulerian) velocity field.

Before proceeding to a more detailed analysis of systems of {\em self-steering} squirmers, we
briefly discuss as reference case the dynamics of pushers without self-steering, i.e., squirmers 
with $\Omega = 0$. 
Interestingly, we do not observe any significant collective behavior without self-steering for 
${\rm Pe} = 128$ as well as $256$, with $\beta = -3$ and $\beta=-5$, in our simulations (see SM, Fig.~S1). It seems 
that thermal fluctuations and consequently rotational diffusion of the microswimmers suppress any structure 
formation at the considered small packing fraction, compare also Ref.~\cite{gasc:23} for a related lattice Boltzmann simulation study of pushers. 
Moreover, the energy spectrum of squirmers strongly deviates from the fluid energy spectrum for large and intermediate $k$ (Fig.~S1), indicating 
that the fluid flows only mildly affect the squirmer dynamics.
As also pointed out in 
Ref.~\cite{dres:11}, presumably even higher densities of force-dipoles are necessary to promote the 
emergence of collective motion engaging both the fluid and microswimmers of pushers without self-steering, 
solely based on hydrodynamic alignment. Indeed, the number density explored in Refs.~\cite{sten:17,bard:19} 
is typically one or two orders of magnitude larger than in our case due to different geometry of active 
particles and their use of force dipoles.

\begin{figure*}
	\centering
  	\includegraphics[width= 2\columnwidth]{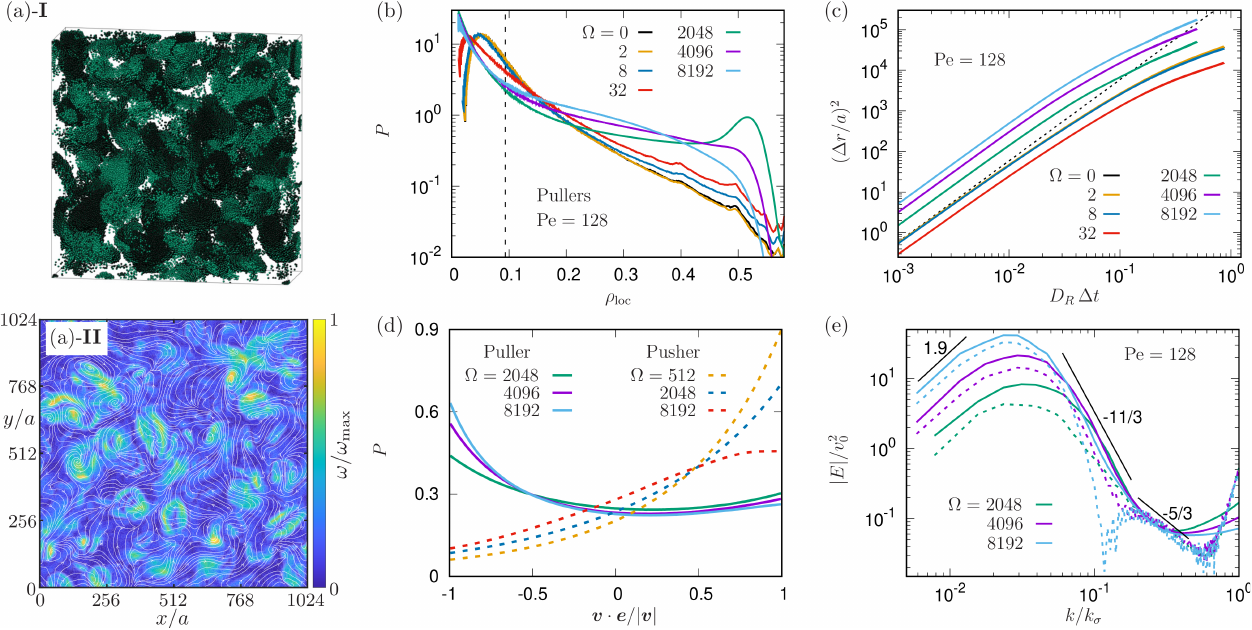}
	\caption{\label{fig:puller}
	{\bf Swarming dynamics in puller systems.}
	(a)-{\bf I} Snapshot of squirmers with their orientation indicated by a petrol hemisphere.
    (a)-{\bf II} Two-dimensional projection of the streamlines of the velocity field (white lines) 
    and the magnitude of the vorticity field (heat map). The parameters are ${\rm Pe}=128$ and $\Omega=8192$.
	(b) Squirmer density distribution  as a function of the local packing fraction for various $\Omega$.
	(c)  Mean-square displacement as a function of time for various $\Omega$ (legend).  The black dashed line indicates the ballistic dynamics of non-interacting squirmers, $(\Delta r)^2 = v_0^2 (\Delta t)^2$.
        (d) Alignment distribution as a function of scalar product of a squirmer's orientation and velocity vector.
	Data for pullers (solid lines) as well as pusher (dashed line) are displayed for comparison.
    (e) Energy spectrum as a function of the wave number $k$ for various $\Omega$ (legend). Energy spectra of fluids and squirmers are represented by solid and dashed lines, respectively. Various power-laws are indicated by short black lines.
	In all figures ${\rm Pe}=128$.
}
\end{figure*}

In sharp contrast, systems of self-steering squirmers display pronounced self-organization, as 
demonstrated in Fig.~\ref{fig:pusher}.
With increasing maneuverability $\Omega$, a scaling regime emerges in the fluid power spectrum for 
wave numbers $k/k_\sigma \lesssim 0.3$, see Fig.~\ref{fig:pusher}(d). Corresponding energy spectra 
of the squirmer motion exhibit the same scaling behavior with the same exponents -- with scaling extending 
even to lower wave numbers $k/k_\sigma \approx 0.02$ for $\Omega=8192$ -- confirming the emergence 
of active turbulence, where the dynamics of squirmers and fluid are strongly correlated on larger 
length scales. The exponents $\nu$ of the energy spectrum, $E(k) \sim (k/k_{\sigma})^{-\nu}$ are 
found to be non-universal, depending on and increasing with $\Omega$, roughly in the range 
$2.8 \lesssim \nu \lesssim 4.0$ for $4 \leq {\rm Pe}/\Omega \leq 64$, see Fig.~S2. Moreover, with increase $\Omega$, not only $\nu$ increases, but also 
the scaling regimes extends to large length scales (smaller $k$, as more squirmers participate in 
the self-organized vortex structures, and the peak height in the energy spectrum increases (up to $|E| \approx 200 v_0^2$ for $\Omega=8192$). This indicates that squirmers attain much higher velocities, 
which is quantitatively confirmed by the mean-square displacement (MSD), see Fig.~\ref{fig:MSD}. 
In the ballistic regime, the MSD $ \langle (\Delta r)^2 \rangle \sim v^2 t^t$, hence, the increasing 
amplitude in Fig.~\ref{fig:MSD} is a direct measure of the increasing speed as $\Omega$ increases 
for $\Omega \geq 512$, i.e., in the regime where a broad scaling region can be identified in the 
energy spectrum (Fig.~\ref{fig:pusher}(d)). 
The P{\'e}clet number ${\rm Pe}$ merely affects turbulent dynamics on large length scales, but the ratio ${\rm Pe}/\Omega$ determines the scaling exponent $\nu$ as shown in Fig.~S2.
We also probe density effects varying the number of squirmers. We first notice that the fluid energy spectra seem to converge for high densities, as shown in Fig.~\ref{fig:pusher}(e), in accordance with the previous report~\cite{bard:19}.
As the packing fraction $\rho$ decreases from $0.186$ to $0.012$ corresponding to 
$\rho_a = 11.9$ and $0.768$, downward shifts in the fluid energy spectra are observed, indicating that fluid stirring by squirmers are not strong at low densities. Consequently, the scaling regime in the energy spectra of squirmers shrinks as the density decreases, e.g., to a narrow range of $0.2 \lesssim k/k_\sigma \lesssim 0.4$ at $\rho = 0.012$.

For $k/k_{\sigma} \gtrsim 0.5$, or length scales smaller than about $2\sigma$, differences between 
the fluid and squirmer energy spectra appear. In this regime, near-field hydrodynamic flows play a significant role. Moreover, the properties of the 
energy spectra depend on ${\rm Pe}$, indicating that noise effects are significant at these small length scales, see Fig.~S3 for fluid thermal energy spectra~\cite{huan:12}.
Therefore, in wet systems and small densities, a strong self-steering of active particles is crucial for the emergence of a large-scale coherent collective motion. Otherwise disorder and hydrodynamic instabilities on small length-scales may prevail~\cite{simh:02}. Such an
observation should also apply to systems where the alignment is mediated via steric repulsion of elongated 
body shapes~\cite{qi:22} or strong hydrodynamic force dipoles~\cite{bard:19}. In any cases, in 
active turbulence of wet systems, collective fluid flows induced by microswimmers build up the collective behavior of microswimmers on large length-scales with fast dynamics. 
In sharp contrast, active turbulence in dry systems~\cite{wens:12.1,keta:24} requires densely packed active particles, because a speedup mechanism as in wet systems is lacking as steric repulsions may only result in a slow-down of active particles. Instead, in dry active turbulence, the scaling regime develops in large $k$, or equivalently, small $|E|$ regimes due to chaotic interparticle collisions on small length scales.



\begin{figure*}
	\centering
  	\includegraphics[width= 2\columnwidth]{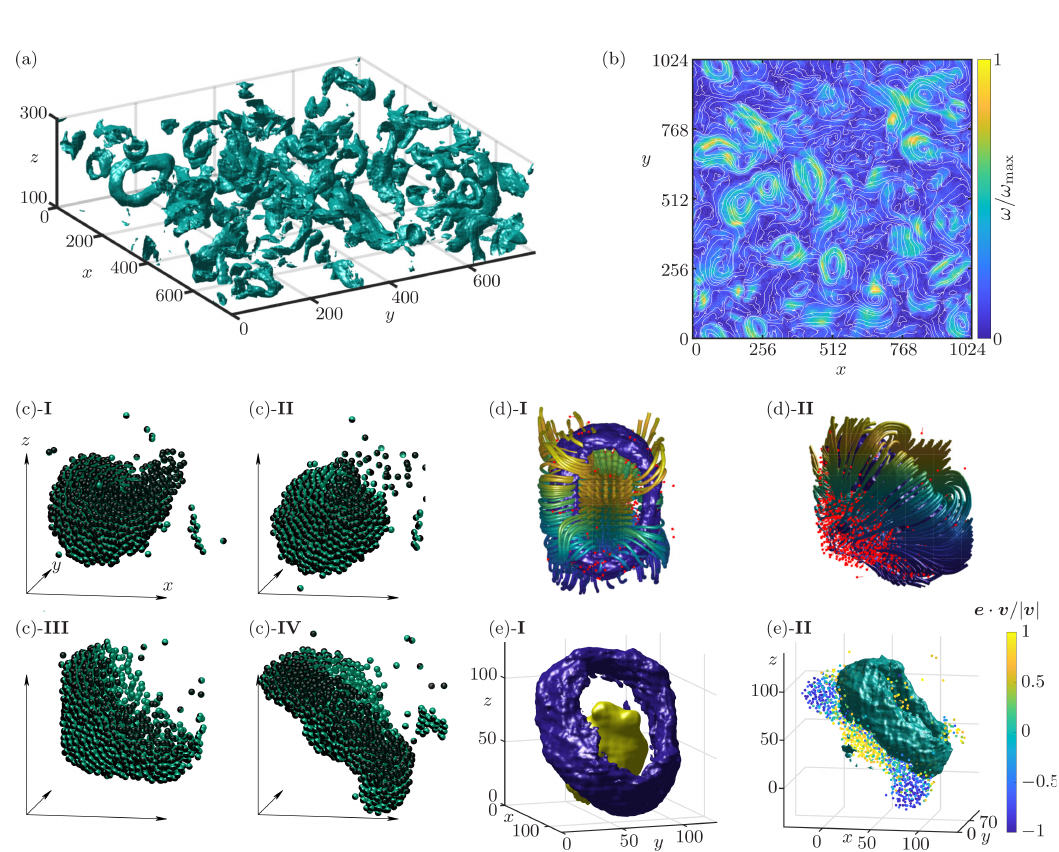}
	\caption{\label{fig:vortex_ring}
	{\bf Formation of vortex rings.}
    (a) System cutout ($100 \leq z/a \leq 300$) of the surface plot for the vorticity field with ${\rm Pe}=128$, $\Omega = 2048$,  $L/a = 768$, and $\omega \approx 0.33 \omega_{\rm max}$, where $\omega_{\rm max}$ indicates the maximum value of $\omega$ in the system at this particular time step.
    (b) Quasi two-dimensional slice of the vorticity field, projected onto the corresponding two dimensions (white lines), together with the magnitude of the vorticity field (heat maps) for ${\rm Pe}=128, \Omega = 8192$, and $L/a = 1024$.
    (c) Time evolution of squirmer configurations during the emergence and destruction of a vortex ring
        for {\bf I} $t = t_0$ (formation of cluster), {\bf II} $t_1 = t_0 + 0.041/D_R$ (formation 
        of jet flow), {\bf III} $t_2 = t_0 + 0.10/D_R$ (emergence of vortex ring), and 
        {\bf IV} $t_3 = t_0 + 0.13/D_R$ (jelly-fish like spreading of squirmers).
    (d) Vortex ring (blue torus, $\omega \approx 0.4 \omega_{\rm max}$) and fluid velocity field 
        (tubes) together with the squirmer positions (red bullets) and orientations (indicated by red bars) at $t=t_2$.
        {\bf I} Top view  and 
        {\bf II}  side-view of the vortex ring, which depict the cluster of squirmers on the bottom-left pulling 
        fluid on the top-right toward the cluster. Squirmers in the frontal region of the cluster are 
        moving along the outer surface of the vortex ring, see also (c)-{\bf III} to identify the 
        locations of the vortex ring and squirmers. 
	(e)-{\bf I} Surface plot of the jet flow ($|{\bm v}_{\rm fl}| \approx 1.6 v_0$) generated at  
        $t=t_1$, together with the vortex ring subsequently formed at $t=t_2$.
        Note that the jet flow is formed at the frontal region of the aligned cluster ((c)-{\bf II}), 
        and directed toward the cluster.
    (e)-{\bf II} Positions and velocities of the squirmers (bullets) at $t=t_3$ ((c)-{\bf IV}), together 
        with the vortex ring previously formed at $t=t_2$. The colors of the squirmers indicate the 
        inner product of their orientation and propulsion direction.
	In (c), (d), and (e), ${\rm Pe}=128$, $\Omega= 2048$.
}
\end{figure*}

\subsection{Pullers: Swarming Dynamics via Self-Steering}

In a systems of self-steering pullers, a rich swarming dynamics develops, as shown in Fig.~\ref{fig:puller}.
The self-organization is characterized by the formation of morphologically complex clusters of 
microswimmers, which, on larger length scales, exhibit visually chaotic movements and exchange 
constitutive squirmers with each others, see Fig.~\ref{fig:puller}(a)-{\bf I} (also Movie S2). 
Still, the puller system exhibits a velocity field with vortical structures 
(see Fig.~\ref{fig:puller}(a)-{\bf II}), surprisingly similar to pushers. 

\begin{figure*}
	\centering
        \includegraphics[width= 2\columnwidth]{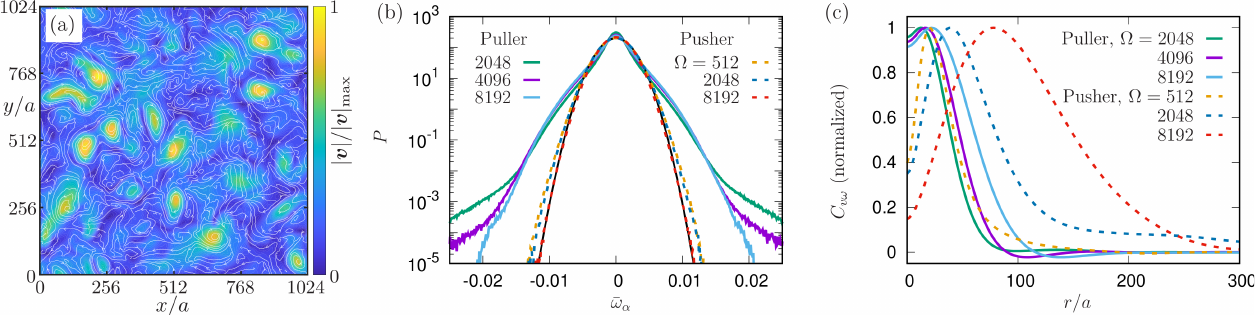}
	\caption{\label{fig:vel_vol}
	{\bf Velocity-vorticity coupling.}
    (a) Vorticity field of pullers (white lines) combined with the magnitude of the velocity field (heat maps) (see Fig.~\ref{fig:vortex_ring}(b)).
        (b) Distribution of the Cartesian components of the vorticity field $\bar{\omega}_\alpha$ as a function of the vorticity $\omega$. The black solid line shows the Gaussian distribution.
        (c) Cross correlation function between the velocity and the vorticity field as a function of the radial squirmer distance.
}
\end{figure*}

\paragraph{Density Modulation.}
The local density distribution is calculated by employing a Voronoi tessellation~\cite{rycr:09}. 
In sharp contrast to pusher systems (Fig.~\ref{fig:pusher}(c)), puller systems exhibit significant density modulations, see Fig.~\ref{fig:puller}(b). For large $\Omega$, $\Omega \gtrsim 4096$, the density distribution is broad, while for $\Omega = 2048$, a clear tendency of segregation is observed (Movie S3) with a low-density peak at $\rho_{\rm loc} \approx 0$, and a high-density 
peak at $\rho_{\rm loc}\approx 0.5$. For small $\Omega$ in the range $128 \le \Omega \le 512$, 
non-mobile clusters appear (see Fig.~S4). 
We focus here on the dynamic clusters, as formation of a dense static cluster may be affected by
a depletion of MPC fluid particles inside the cluster -- which is related to the (weak) compressibility 
of the MPC fluid, and artificially enhance cluster stability~\cite{thee:18}. For $\Omega \leq 32$, 
unimodal distributions are recovered, but with a peak at a density smaller than the 
global density of $\rho = 0.093$, and with fatter tails than those in pusher systems, which reflects 
a clustering tendency reported for pullers~\cite{thee:18,zant:22}.

\paragraph{Decoupling of Puller Orientation and Velocity.}
The emergence of a high-density regime for $\Omega \ge 2048$ needs to be distinguished from 
motility-induced phase separation (MIPS) in dry systems. First, squirmers in a (small) cluster exhibit 
local polar order, giving rise to a coherent directional motion of the cluster. Second, the 
pullers actually swim faster on average than their bare self-propulsion speed $v_0$, as shown in 
Fig.~\ref{fig:puller}(c) -- as in the system of pushers (see Fig.~\ref{fig:pusher}(e)). 
In particular, decoupling of self-propulsion and velocity of squirmers is far more significant 
than in MIPS. Even situations, where pullers are driven backward occur frequently (Movie S4). Indeed, 
as shown in Fig.~\ref{fig:puller}(d), the probability that the velocity of a squirmer is 
anti-parallel to its orientation, i.e., ${\bm v}/|{\bm v}| = -{\bm e}$, is even higher than 
for the parallel case, i.e., ${\bm v}/|{\bm v}| = {\bm e}$. Therefore, we conclude that 
hydrodynamic interactions between self-steering pullers dominate over the self-propulsion forces. 
The attractive hydrodynamic interactions between aligned pullers in a head-to-tail
configuration promotes the formation of dense clusters. Yet, as we will demonstrate below, 
dense clusters are not static, but exhibit a highly dynamical morphology.

\paragraph{Decoupling of Puller and Fluid Dynamics.} 
The kinetic energy spectra $E(k)$ for squirmer motion and the fluid in puller systems are presented in Fig.~\ref{fig:MSD}. 
While for the swarming of pullers, $E(k)$ exhibits a power-law decay in the intermediate wave-number 
regime ($0.04 < k/k_{\sigma} < 0.2$) for the fluid, as in pusher systems, several features 
significantly deviate from those of pushers, which again demonstrates the uniqueness of puller 
swarming. Most of all, a pronounced mismatch between squirmer and fluid energy spectra is observed 
in the intermediate regime of $k/k_\sigma \approx 0.1$ (see Fig.~S5(a) and Fig.~S7(a) for the corresponding correlation functions), which indicates that pullers are not simply 
driven by the fluid flow. 
The exponents of the three scaling regimes in Fig.~\ref{fig:puller}(e) show universal behavior for the 
investigated range of $\Omega$. Notably, in the case of the squirmer energy spectra for 
$0.2 < k/k_{\sigma} < 0.6$, which correspond the length scales of the high-density intra-cluster regime 
where steric repulsion and near-field hydrodynamic interactions dominate, the obtained value of 
the exponent $\nu \simeq 5/3$ is in accordance with the Kolmogorov exponent for classical 
hydrodynamic turbulence and with active turbulence of dense quasi-two-dimensional pusher systems 
\cite{qi:22}. For larger length scales, an exponent $\nu\simeq 11/3$ is obtained, which persists under variation of global densities, see Fig.~S6. A similar value 
has been found previously in Lattice Boltzmann simulations of extended force dipoles~\cite{bard:19}, although at higher densities and for pushers. 

\paragraph{Formation of vortex ring.}
An even more striking feature is observed in configurations of the vorticity field. 
Visual observation of the time evolution of the system indicates a typical dynamical behavior in the morphology of clusters, which involves the pulsatile transformation of aggregates from a spherical shape into jellyfish-like 
arrangements. 
In terms of fluid mechanics, 
the jellyfish-like morphology suggests formation of a vortex ring,
which is indeed confirmed by emergence of toroidal structures in the vorticity fields extracted from the simulations, see Fig.~\ref{fig:vortex_ring}(a). As shown in Fig.~\ref{fig:vortex_ring}(b), the vorticity field exhibits whirling patterns within regions where the magnitude of the vorticity field is large. 

For a detailed illustration, we consider the small system size $L/a = 128$, where only a single cluster emerges (Movie S5). As shown in Fig.~\ref{fig:vortex_ring}(c)-{\bf I}, the dynamics 
initiates with formation of a cluster. Then, due to alignment, squirmers rotate, and a polar order 
emerges within the cluster, see Fig.~\ref{fig:vortex_ring}(c)-{\bf II}, and also Fig.~S5(b) for the corresponding order parameter. Such an ordered structure gives rise to 
a strong collective fluid flow, which generates a pronounced jet in front of the cluster, see the yellow 
surface in Fig.~\ref{fig:vortex_ring}(e)-{\bf I}. Notably, the jet flow is self-generated via active 
stirring of microswimmers in this case, instead of external perturbation as in passive
hydrodynamic fluids. 
Subsequently, a spread-out motion of squirmers is initiated (Fig.~\ref{fig:vortex_ring}(c)-{\bf III}). 
Simultaneously, a vortex ring is formed around the cluster (blue ring in Fig.~\ref{fig:vortex_ring}(e)-{\bf I}), 
while squirmers are moving forward. As shown in Fig.~\ref{fig:vortex_ring}, the velocity field is indeed wrapping around the vortex ring, in accordance with Fig.~\ref{fig:vortex_ring}(b).
Then, the pullers continue to spread out, rolling about the 
region where the vortex ring forms, see Fig.~\ref{fig:vortex_ring}(c)-IV. While swimmers at the cluster 
center swim forward, they are dragged backward at the periphery, as shown in 
Fig.~\ref{fig:vortex_ring}(e)-{\bf II}, which contributes to the anomalous behavior in the distribution 
of ${\bm v}\cdot {\bm e}/|{\bm v}|$ (see Fig.~\ref{fig:puller}(d)). Eventually, the cluster dissolves, 
ending its life cycle.


\subsection{Velocity-Vorticity Coupling}


The sequential time evolution described so far indicates a strong coupling between the fluid velocity and the vorticity field in a puller cluster. As shown in Fig.~\ref{fig:vel_vol}(a), the rotation of the vorticity field for pullers is indeed centered at regions with a strong velocity field.

For a more quantitative characterization, we examine the distribution of the Cartesian components of the velocity and vorticity fields. In pusher systems, both the velocity and vorticity fields (Fig.~S7(b) and Fig.~\ref{fig:vel_vol}(b), respectively) exhibit a Gaussian distribution, which is an indicator of active turbulence~\cite{qi:22}. For pullers, both deviate from a Gaussian, demonstrating that the swarming dynamics of pullers is not active turbulence.
Specifically, the velocity distribution of pullers exhibits ``fat" 
exponential tails, as shown in Fig.~S7(b), in line with the emergence of stronger fluid flows than ``expected", i.e., the occurrence of jet plumes induced by aligned pullers. 
Also the vorticity field for pullers shows a broader distribution than that of pushers as displayed in Fig.~\ref{fig:vel_vol}(b). Furthermore, the vorticity distribution for pullers in Fig.~\ref{fig:vel_vol}(b) shows a rather sharp peak at $\bar{\omega}_{\alpha}$, the average of three Cartesian components $\omega_\alpha$ ($\alpha=x,y,z$) of $\boldsymbol{\omega}$, indicating a weak separation between regions with strong and weak vorticity. 


A more fundamental difference in velocity-vorticity coupling is revealed by a cross correlations between the magnitude of 
vorticity and that of the velocity field ($|\omega (\bm r)|$ and $|{\bm v} (\bm r)|$), as defined in 
Eq.~\eqref{eq:cross}. In Fig.~\ref{fig:vel_vol}(c), the various cross 
correlations for pushers exhibit a pronounced peak at $r/a=20$, $60$, and $120$ for $\Omega=512$, $2048$, 
and $8192$, respectively, while the cross correlation at $r=0$ is not strong. Hence, for pushers, 
the velocity field of a vortex is weak at the center but strong at intermediate regions between the 
center and periphery. In sharp contrast, for pullers, the cross correlation between vorticity 
and velocity fields is already strong at small distances, which demonstrates that a strong velocity 
field generates a strong vorticity in the immediate vicinity of the jet flow. Moreover, the cross 
correlation decays faster than for pushers, assuming negative values before approaching 
zero.

\section{Discussion and Conclusions}

We have studied self-organization and dynamics in three-dimensional wet systems of self-steering 
squirmers, which aim for alignment of their orientation with that of their neighbors. We demonstrate 
that alignment via hydrodynamic self-steering gives rise to a rich collective behavior in such polar 
active fluids, depending on the type of active stresses, i.e., whether the microswimmers are pushers or 
pullers. In both cases, an essential role of hydrodynamics is the breaking of long-range polar order, 
which causes the emergence of chaotic behavior.

For pushers, the particle distribution is quite homogeneous, the distribution of the Cartesian velocity
components is Gaussian, and the kinetic energy spectrum displays a peak and subsequent power-law decay 
with increasing wave vector, which indicates active-turbulent behavior. An intriguing feature is that
strong self-steering enhances the coherent movement of microswimmers, which leads to collective particle
motion with speeds much faster then the individual swim speed -- as can be seen in the increasing 
magnitude of the peak in the energy spectrum, combined with an extension of the scaling regime toward 
large length scales. This implies that large-scale flows are induced by the collective motion, which 
drag the microswimmers along  and supersede their individual motion. Thus, the polarity field and the 
fluid flow field are strongly coupled.

For pullers, another type of self-organization emerges, which is strictly distinguished from 
motility-induced phase separation (MIPS) of dry ABP systems as well as  active turbulence of pushers.
The particle density is now found to be very inhomogeneous, as the pullers tend to form clusters.
However, these clusters are not static, but appear as quite unstable. The particle alignment
inside the cluster generates a strong fluid jet and a vortex ring, which pulls apart the cluster and
leads to its disintegration. These strong flows imply that the probability of fast fluid flows is enhanced,
which is reflected in the emergence of fat tails in the velocity distribution.

In contrast, wet systems of self-propelled squirmers {\em without} self-steering display no 
interesting collective behavior in three dimensions, not even at high squirmer volume fractions. 
 
Our numerical observations for ensembles of self-steering pullers challenges current theoretical 
views on collective behaviors in wet active systems. So far, it has been typically assumed that 
the polarity and velocity fields of active fluids are essentially identical, based on the 
assumption of a nearly homogeneous distribution of active particles. Heterogeneous
densities have been observed recently in  models of compressible polar active fluids for bacterial suspensions \cite{worl:21,worl:21.1}; however, the mechanism is entirely different in this case,
as hydrodynamic interactions are not considered, and clustering is driven by a strong dependence of
self-propulsion speed on the local density.

Our results demonstrate that in wet systems of self-steering microswimmers in three dimensions, 
the interplay of the particle density and polarity, and of the fluid velocity field can 
give rise to a surprisingly rich variety of emergent behaviors -- already for a highly simplified
model systems with only a single particle type.

\section{Materials and Methods}

\subsection{Mesoscale Fluid Model: Multi-Particle Collision Dynamics}

We adopt the multiparticle collision dynamics (MPC) method~\cite{kapr:08,gomp:09}, a particle-based mesoscale simulation approach, as model for the fluid. Specifically, we employ the stochastic rotation variant of MPC with angular momentum conservation (MPC-SRD+a)~\cite{nogu:08} and the cell-level Maxwell-Boltzmann scaling thermostat~\cite{huan:10.1}. The algorithm consists of alternating streaming and collision steps. In the streaming step, the MPC point particles of mass $m$ propagate ballistically over the collision time interval $h$, denoted as collision time. In the collision step, ﬂuid particles are sorted into the cells of a cubic lattice of lattice constant $a$ deﬁning the collision environment. Then, their relative velocities, with respect to the center-of-mass velocity of the collision cell, are rotated around a randomly oriented axes by a ﬁxed angle $\alpha$. The algorithm conserves mass, linear, and angular momentum on the collision-cell level, while thermal fluctuations are automatically incorporated. More details are described in Refs.~\cite{thee:16.1,qi:22} and the SM, which refers to Refs.~\cite{omel:98,lamu:01,padd:05,thee:14} additionally. 
Our GPU-based, highly parallelized implementation employed for the simulations is proposed in Ref.~\cite{west:24}.

We use the average MPC fluid density (particles per collision cell) $\langle N_c \rangle = 20$, the collision time $h=0.02 a\sqrt{m/(k_{\rm B}T)}$ and the rotation angle $\alpha = 130^{\circ}$, which yield the fluid viscosity of $\eta = 42.6\sqrt{mk_{\rm B}T}/a^2$~\cite{nogu:08,thee:15}. With the squirmer radius $R_{\rm c}= 3 a$, these MPC parameters yield the rotational diffusion coefficient $D_R = 4.1 \times 10^{-5}\sqrt{k_{\rm B}T/m}/a$.

\subsection{Hydrodynamic Self-Steering}

Hydrodynamic self-steering of squirmers is achieved via adaptive surface flow fields, given as 
\begin{align}
u_\phi =& \frac{3}{2}v_0\sin{\theta} (1+\beta \cos{\theta}) - \frac{1}{R_{\rm sq}^2}(\tilde{C}_{11}\cos{\phi} - C_{11}\sin{\phi}) \label{eq:uph} \\
u_\theta =&\frac{\cos{\theta}}{R_{\rm sq}^2} (C_{11}\cos{\phi} +\tilde{C}_{11}\sin{\phi}), 
\label{eq:uth}
\end{align}
where $\theta$ and $\phi$ are the polar and azimuthal angles in a body-fixed reference frame. The parameter $v_0$ characterizes the swim speed and $\beta$ the strength of the force dipole, where $\beta <0$ for pullers and $\beta>0$ for pushers~\cite{elge:15,shae:20}. $C_{11}$ and $\tilde{C}_{11}$ control the magnitudes of non-axisymmetric surface-flow components, leading to rotational motion of the body~\cite{pak:14,goh:23}. Specifically, the non-axisymmetric flow fields enable the adaptive motion of Eq.~\eqref{eq:omegai} via
\begin{align}
C_{11} &= C_0 R_{\rm sq}^3 ({\bm e} \times {\bm e}_{\rm aim})\cdot {\bm e}_x, \\
\tilde{C}_{11} &= C_0 R_{\rm sq}^3 ({\bm e} \times {\bm e}_{\rm aim})\cdot {\bm e}_y,
\end{align}
where ${\bm e}_x$ and ${\bm e}_y$ are the unit vectors along the axis of the body-fixed reference frame. For the self-propulsion, we use $v_0/\sqrt{k_{\rm B}T/m} = 0.007872$ and $0.031488$, which correspond to ${\rm Pe} = v_0/(\sigma D_R) =32$ and $128$, and ${\rm Re}=0.022$ and $0.089$, respectively. The values of the self-steering strength $C_0$ are varied from $0$ to $0.335872 \sqrt{k_{\rm B}T/m}/a$ yielding $0 \leq \Omega \leq 8192$.

\subsection{Steric squirmer interaction}

Steric repulsion between two squirmers is described by the separation-shifted Lennard-Jones potential 
\begin{align}
U_{\rm LJ} (d_s) = 4\epsilon_0 \left[ 
\left( \frac{\sigma_0}{d_s +\sigma_0} \right)^{12}
-\left( \frac{\sigma_0}{d_s +\sigma_0} \right)^{6}
+\frac{1}{4} \right],
\end{align}
for $d_s < (2^{1/6}-1)\sigma_0$ and zero otherwise, where $d_s$ indicates the surface-to-surface distance between the two squirmer. To avoid loss of hydrodynamic interactions when two squirmers contact with each other, we also include a virtual safety distance $d_v$~\cite{thee:16,qi:22}, which leads to the effective distance $d_s = r_c - \sigma -2d_v$, where $r_c$ denotes the center-to-center distance and $\sigma$ is the squirmer diameter. We choose $\sigma_0 = 2d_v$. Numerically, the equations of motion for the rigid-body dynamics of the squirmers are solved by the velocity-Verlet algorithm, see SM for more details.

\subsection{Spatial correlation and Energy Spectrum}

\paragraph{Squirmers: Particle-based approach.} 
The spatial velocity correlation function of the squirmers is defined as
\begin{align}
C_{\rm sq} (r) = \frac{1}{\bar{v}^2}\frac{\left\langle \sum_{i\neq j} {\bm v}_i\cdot {\bm v}_j \delta (r - |{\bm r}_i -{\bm r}_j|) \right\rangle}{\left\langle \sum_{i\neq j} \delta (r - |{\bm r}_i -{\bm r}_j|) \right\rangle}.
\end{align}
where $\bar{v}^2 \equiv \left( \sum_i |{\bm v}_i|^2 /N \right)$.
Here, we use the velocity averaged over a short time interval $\Delta t$, for which we consider $\Delta t = 0.26  \sigma/v_0$. The energy spectrum can be calculated via Fourier transformation. Here we consider the Fourier sine transform~\cite{batc:59}
\begin{align}
E_{\rm sq}(k) = \frac{k}{\pi} \int {\rm d}r\,r \sin{kr}\, \bar{v}^2 C_{\rm sq}(r).
\end{align}

\paragraph{Fluid: Field-based approach.} 
We first extract the fluid velocity field ${\bm v}_{\rm fl}$ from the simulation data by introducing a grid dividing the whole system into $N_g^3$ cells. The velocities of all MPC particles, averaged over a short time interval $\Delta t$ as for squirmers, are additionally averaged over each cell to obtain ${\bm v}_{\bm n}$ with ${\bm n}=(n_x,n_y,n_z)^T$ for $n_i=0,\ldots, N_g-1$ and $i\in \{x,y,z\}$. Then the discrete Fourier transform 
\begin{align} \label{eq:fourier}
{\bm v}_{\rm fl} ({\bm k}) = \frac{1}{N_g^3}\sum_{\bm n} {\bm v}_{\rm sq}({\bm n}) e^{i2\pi a {\bm k}\cdot {\bm n}/N_g}
\end{align}
is performed.
The energy spectrum is calculated straightforwardly via~\cite{liu:21}
\begin{align} \label{eq:ek_vk}
E_{\rm fl}({\bm k})= \frac{1}{2} |{\bm v}_{\rm fl}({\bm k})|^2,
\end{align}
which is then averaged over all directions of $\bm k$ to obtain $E_{\rm fl}(k)$. Then, the spatial velocity correlation function is obtained via the Fourier transformation 
\begin{align} \label{eq:inv_fourier}
C_v ({\bm n}) = \sum_{\bm n} \left\langle E_{\rm fl}({\bm k}) \right\rangle e^{-i2\pi a {\bm k}\cdot {\bm n} /N_g},
\end{align}
from which we calculate $C_{\rm fl}(r)$ by averaging over all directions of $\bm n$.
To reduce noise effects at small length scales, velocity fields are averaged over boxes with side lengths $\sigma$, $2\sigma$, or $3\sigma$. 

The vorticity field is then defined as 
\begin{align} \label{eq:vorticity}
{\bm \omega} ({\bm n}) \equiv {\bm \nabla}_{\bm n} \times {\bm v}_{\rm fl} ({\bm n}).
\end{align}
Numerically, the vorticity field is calculated by the five-point stencil method from the velocity field. The vorticity spatial correlation $C_\omega$ are also obtained from Eqs.~\eqref{eq:fourier}-\eqref{eq:inv_fourier}.
Moreover, a gliding time average is performed for the velocity with a time window $\Delta t$, corresponding to $v_0 \Delta t \approx 0.26\sigma$.

\paragraph{Cross correlation.}
We again utilize Fourier transformation to calculate cross correlations between velocity and vorticity fields. Specifically, we first calculate the magnitudes of velocity and vorticity fields, which are then shifted by their average values, i.e., $\tilde{v}(\bm n) \equiv v(\bm n) - \sum_{\bm n} v (\bm n)/N_g^3$ and $\tilde{\omega}(\bm n) \equiv \omega(\bm n) - \sum_{\bm n} \omega (\bm n)/N_g^3$. Then, from the Fourier transforms of the fields $\tilde{v} ({\bm k})$ and $\tilde{\omega} ({\bm k})$, the cross correlation is obtained via
\begin{align} \label{eq:cross}
C_{v\omega}({\bm n}) = \sum_{\bm n} \left\langle \tilde{v}({\bm k}) \tilde{\omega}^* ({\bm k}) \right\rangle e^{-i2\pi a {\bm k}\cdot{\bm n}/N_g},
\end{align}
where superscript $*$ indicates the complex conjugate. $C_{v\omega}(r)$ is obtained by averaging over all $\bm n$ directions.

\section*{Acknowledgements}
\noindent
{\bf Funding:} The authors gratefully acknowledge the Gauss Centre for Supercomputing e.V. (www.gauss-centre.eu) for funding this project by providing computing time through the John von Neumann Institute for Computing (NIC) on the GCS Supercomputer JUWELS at J\"ulich Supercomputing Centre (JSC). {\bf Author contributions:} G.G. and R.G.W. designed the research. E.W. wrote the simulation code. S.G. performed the simulations and analyzed the data. S.G., R.G.W., and G.G. discussed the results and wrote the manuscript together. {\bf Competing interests:} The authors declare that they have no competing interests. {\bf Data and materials availability:} All data needed to evaluate the conclusions in the paper are available from the corresponding author upon reasonable request.

\end{document}